\documentclass[aps,prl,amsmath,amssymb,reprint,superscriptaddress, nofootinbib,
]{revtex4-1}

\usepackage{hepunits}
\usepackage{graphicx}\usepackage{dcolumn}\usepackage{bm}\usepackage[mathlines]{lineno}\usepackage{lipsum}
\usepackage{xspace}
\usepackage[nolist]{acronym}
\usepackage{color}
\usepackage[caption=false]{subfig}
\usepackage{circuitikz}
\usepackage{comment}
\usepackage[normalem]{ulem}

\graphicspath{{images/}}

\usepackage{etoolbox}
\makeatletter
\patchcmd\linenumberpar{\@LN@parpgbrk}{\penalty\@LN@parpgpen\relax}{}{}
\makeatother

\begin{document}

\title[ABRACADABRA-10 cm]{First Results from ABRACADABRA-10\,cm: A Search for Sub-$\bm{\mu}$eV\\Axion Dark Matter}
\author{Jonathan~L.~Ouellet}
\email{ouelletj@mit.edu}
\affiliation{Laboratory for Nuclear Science, Massachusetts Institute of Technology, Cambridge, MA 02139, U.S.A.}

\author{Chiara~P.~Salemi}
\affiliation{Laboratory for Nuclear Science, Massachusetts Institute of Technology, Cambridge, MA 02139, U.S.A.}

\author{Joshua~W.~Foster}
\affiliation{Leinweber Center for Theoretical Physics, Department of Physics, University of Michigan, Ann Arbor, MI 48109, U.S.A.}

\author{Reyco~Henning}
\affiliation{University of North Carolina, Chapel Hill, NC 27599, U.S.A.}
\affiliation{Triangle Universities Nuclear Laboratory, Durham, NC 27708, U.S.A.}

\author{Zachary~Bogorad}
\affiliation{Laboratory for Nuclear Science, Massachusetts Institute of Technology, Cambridge, MA 02139, U.S.A.}

\author{Janet~M.~Conrad}
\affiliation{Laboratory for Nuclear Science, Massachusetts Institute of Technology, Cambridge, MA 02139, U.S.A.}

\author{Joseph~A.~Formaggio}
\affiliation{Laboratory for Nuclear Science, Massachusetts Institute of Technology, Cambridge, MA 02139, U.S.A.}

\author{Yonatan~Kahn}
\affiliation{Princeton University, Princeton, NJ 08544, U.S.A.}
\affiliation{Kavli Institute for Cosmological Physics, University of Chicago, Chicago, IL 60637, U.S.A.}

\author{Joe~Minervini}
\affiliation{Plasma Science and Fusion Center, Massachusetts Institute of Technology, Cambridge, MA 02139, U.S.A.}

\author{Alexey~Radovinsky}
\affiliation{Plasma Science and Fusion Center, Massachusetts Institute of Technology, Cambridge, MA 02139, U.S.A.}

\author{Nicholas~L.~Rodd}
\affiliation{Berkeley Center for Theoretical Physics, University of California, Berkeley, CA 94720, U.S.A.}
\affiliation{Theoretical Physics Group, Lawrence Berkeley National Laboratory, Berkeley, CA 94720, U.S.A.}

\author{Benjamin~R.~Safdi}
\affiliation{Leinweber Center for Theoretical Physics, Department of Physics, University of Michigan, Ann Arbor, MI 48109, U.S.A.}

\author{Jesse~Thaler}
\affiliation{Center for Theoretical Physics, Massachusetts Institute of Technology, Cambridge, MA 02139, U.S.A.}

\author{Daniel~Winklehner}
\affiliation{Laboratory for Nuclear Science, Massachusetts Institute of Technology, Cambridge, MA 02139, U.S.A.}

\author{Lindley~Winslow}
\email{lwinslow@mit.edu}
\affiliation{Laboratory for Nuclear Science, Massachusetts Institute of Technology, Cambridge, MA 02139, U.S.A.}

 \date{March 12, 2019}

\begin{abstract}
The axion is a promising dark matter candidate, which was originally proposed to solve the strong-CP problem in particle physics.
To date, the available parameter space for axion and axion-like particle dark matter is relatively unexplored, particularly at masses $m_a\lesssim1\,\mu$eV. ABRACADABRA is a new experimental program to search for axion dark matter over a broad range of masses, \mbox{$10^{-12}\lesssim m_a\lesssim10^{-6}$\,eV}. ABRACADABRA-10\,cm is a small-scale prototype for a future detector that could be sensitive to the QCD axion. In this Letter, we present the first results from a 1\,month search for axions with ABRACADABRA-10\,cm. We find no evidence for axion-like cosmic dark matter and set 95\% C.L.\ upper limits on the axion-photon coupling between \mbox{$g_{a\gamma\gamma}<1.4\times10^{-10}$\,GeV$^{-1}$} and \mbox{$g_{a\gamma\gamma}<3.3\times10^{-9}$\,GeV$^{-1}$} over the mass range $3.1\times10^{-10}$\,eV -- $8.3\times10^{-9}$\,eV. These results are competitive with the most stringent astrophysical constraints in this mass range.
\end{abstract}

\maketitle

\begin{acronym}
\acro{ABRA}[ABRACADABRA]{{\bf A} {\bf B}roadband/{\bf R}esonant {\bf A}pproach to {\bf C}osmic {\bf A}xion {\bf D}etection with an {\bf A}mplifying {\bf B}-field {\bf R}ing {\bf A}pparatus}
\acro{PSD}{power spectral density}
\acro{ALP}{Axion-like particles}
\acro{DM}{dark matter}
\acro{ADM}{Axion DM}
\acro{DFT}{discrete Fourier transform}
\acro{FLL}{flux-lock feedback loop}
\acro{SNR}{signal-to-noise ratio}
\acro{LEE}{look-elsewhere effect}
\acro{TS}{test-statistic}
\acro{POM}{polyoxymethylene}
\acro{PTFE}{polytetrafluoroethylene}
\acro{MC}{Monte Carlo}
\acro{WIMP}{Weakly Interacting Massive Particle}
\acro{PQ}{Peccei-Quinn}
\end{acronym}

\newcommand{\ABRA}{\ac{ABRA}\xspace}
\newcommand{\PSD}{\ac{PSD}\xspace}
\newcommand{\abra}{ABRACADABRA-10\,cm\xspace}
\newcommand{\ALP}{\ac{ALP}\xspace}
\newcommand{\ALPs}{\ac{ALP}s\xspace}
\newcommand{\DM}{\ac{DM}\xspace}
\newcommand{\ADM}{\ac{ADM}\xspace}
\newcommand{\gagg}{\ensuremath{g_{a\gamma\gamma}}\xspace}
\newcommand{\rhoDM}{\ensuremath{\rho_{\rm DM}}\xspace}
\newcommand{\DFT}{\ac{DFT}\xspace}
\newcommand{\FLL}{\ac{FLL}\xspace}
\newcommand{\SNR}{\ac{SNR}\xspace}
\newcommand{\LEE}{\ac{LEE}\xspace}
\newcommand{\TS}{\ac{TS}\xspace}
\newcommand{\POM}{\ac{POM}\xspace}
\newcommand{\PTFE}{\ac{PTFE}\xspace}
\newcommand{\MC}{\ac{MC}\xspace}
\newcommand{\WIMP}{\ac{WIMP}\xspace}
\newcommand{\PQ}{\ac{PQ}\xspace}

\newcommand{\Px}{\ensuremath{\bar{\mathcal{F}}}}
\newcommand{\Pten}{\ensuremath{\bar{\mathcal{F}}_\mathrm{10M}}\xspace}
\newcommand{\Pone}{\ensuremath{\bar{\mathcal{F}}_\mathrm{1M}}\xspace}
\newcommand{\Phun}{\ensuremath{\mathcal{F}_\mathrm{100k}}\xspace} 
\section{Introduction}
The particle nature of \DM in the Universe remains one of the greatest mysteries of contemporary physics. Axions are an especially promising candidate as they can simultaneously explain both the particle nature of \DM and resolve the strong-CP problem of quantum chromodynamics (QCD) \cite{Peccei1977,Abbott1982,Preskill1983,Dine1983,Weinberg:1977ma,Wilczek1977}. \ALP are generically expected to have a coupling to electromagnetism of the form \cite{Sikivie1983}
\begin{equation}
\mathcal{L} \supset -\frac{1}{4}\gagg a\widetilde{F}_{\mu \nu}F^{\mu \nu} = \gagg a \,\mathbf{E}\cdot \mathbf{B},
\label{eq:gagg}
\end{equation}
where $\gagg$ is the axion-photon coupling. The QCD axion is predicted to have a narrow range of couplings proportional to the axion mass, while a general \ALP may have any \gagg. In this work, ``axion'' refers to a general \ALP. \ADM with mass $m_a\ll1$\,eV behaves today as a classical field oscillating at a frequency $f = m_a/(2\pi)$ \cite{Preskill1983,Dine1983}. The Lagrangian~(\ref{eq:gagg}) implies that a time-dependent background density of \ADM modifies Maxwell's equations. In particular, in the presence of a static magnetic field $\mathbf{B}_0$, \ADM generates an oscillating magnetic field, $\mathbf{B}_a$, as if sourced by an effective AC current density parallel to $\mathbf{B}_0$ \cite{Ouellet2018a},
\begin{equation}
\mathbf{J}_{\rm eff} = \gagg \sqrt{2\rhoDM} \mathbf{B}_0 \cos(m_a t).
\label{eq:Jeff}
\end{equation}
Here $\rhoDM$ is the local \DM density, which we take to be 0.4\,GeV/cm$^3$ \cite{Catena:2009mf,Iocco:2011jz}.
The \ABRA\ experiment, as first proposed in \cite{ABRA2016}, is designed to search for the axion-induced field, $\mathbf{B}_a$, generated by a toroidal magnetic field (see also \cite{Sikivie:2013laa} for a proposal using a solenoidal field). ABRACADABRA searches for an AC magnetic flux through a superconducting pickup loop in the center of a toroidal magnet, which should host no AC flux in the absence of \ADM. The time-averaged magnitude of the flux through the pickup loop due to $\mathbf{B}_a$ can be written as
\begin{equation}
|\Phi_a|^2 = \gagg^2\rhoDM V^2\mathcal{G}^2B_{\rm max}^2\equiv A,
\label{eq:pickup_flux_power}
\end{equation}
where $V$ is the volume of the toroid, $\mathcal{G}$ is a geometric factor calculated for our toroid to be 0.027 \cite{ABRA_10cm_Technical}, and $B_{\rm max}$ is the maximum $B$-field in the toroid. The pickup loop is read out using a SQUID current sensor, where an axion signal would appear as a small-amplitude, narrow ($\Delta f/f\sim 10^{-6}$) peak in the \PSD of the SQUID output at a frequency given by the axion mass. The present design uses a simplified broadband readout, but the same approach can be significantly enhanced using resonant amplification and recent developments in powerful quantum sensors \cite{Chaudhuri:2018rqn,Ahmed:2018oog}, which is the subject of future work. 

In this Letter, we present first results from \abra, probing the axion-photon coupling $\gagg$ for ADM in the frequency range $f~\in~[75\,\rm{kHz}, \ 2\,{\rm MHz}]$, corresponding to axion masses $m_a~\in~[3.1 \times 10^{-10}, \ 8.3 \times 10^{-9}]\ \eV$. This mass range is highly motivated for QCD axions, where the axion decay constant lies near the GUT scale and is easily compatible with pre-inflationary \PQ breaking in a variety of models, including grand unified theories \cite{DiLuzio:2018gqe} or string compactifications \cite{Svrcek:2006yi,Arvanitaki:2009fg}, and such low-mass axions may be favored anthropically \cite{Tegmark:2005dy}. Additionally, such light \ALPs may explain the previously-observed transparency anomaly of the Universe to TeV gamma-rays~\cite{Horns:2012fx,Meyer:2013pny,Roncadelli2008,Burrage2009}, though in this case the \ALP is not required to be DM. Recently, this mass range has gathered significant experimental interest \cite{Sikivie:2013laa,ABRA2016,Silva-Feaver2016,PhysRevX.4.021030,Gatti:2018ojx,DeRocco:2018jwe,Liu:2018icu} to name a few, or see~\cite{Irastorza:2018} for a comprehensive review. Furthermore, this mass range is highly complementary to that probed by the ADMX experiment~\cite{Asztalos2001,Asztalos2009,ADMX2018}, HAYSTAC \cite{HAYSTAC2018a}, and other microwave cavity experiments \cite{PhysRevD.42.1297,PhysRevLett.59.839,PhysRevLett.80.2043}, which probe $m_a \sim 10^{-6}-10^{-5}\,\eV$. Our result represents the most sensitive laboratory search for \ADM below 1\,$\mu$eV, is competitive with leading astrophysical constraints from CAST \cite{CAST2017},  and probes a region where low-mass \ALPs which can accommodate all the \DM of the universe without overclosure \cite{Davoudiasl:2015vba,Graham:2018jyp,Agrawal:2017eqm,2010PhRvD..81f3508V,Co:2016vsi}, as well as particular models of QCD axions with enhanced photon couplings \cite{Agrawal:2017cmd,Farina:2016tgd}. Aside from the ALP models currently being probed, this result is a crucial first step towards a larger-scale version of \ABRA sensitive to the smaller values of \gagg relevant for the typical QCD axion models in the mass range where axions can probe GUT-scale physics.

\section{Magnet and Cryogenic Setup}
\begin{figure}[h!]
  \centering
  \includegraphics[width=0.48\textwidth]{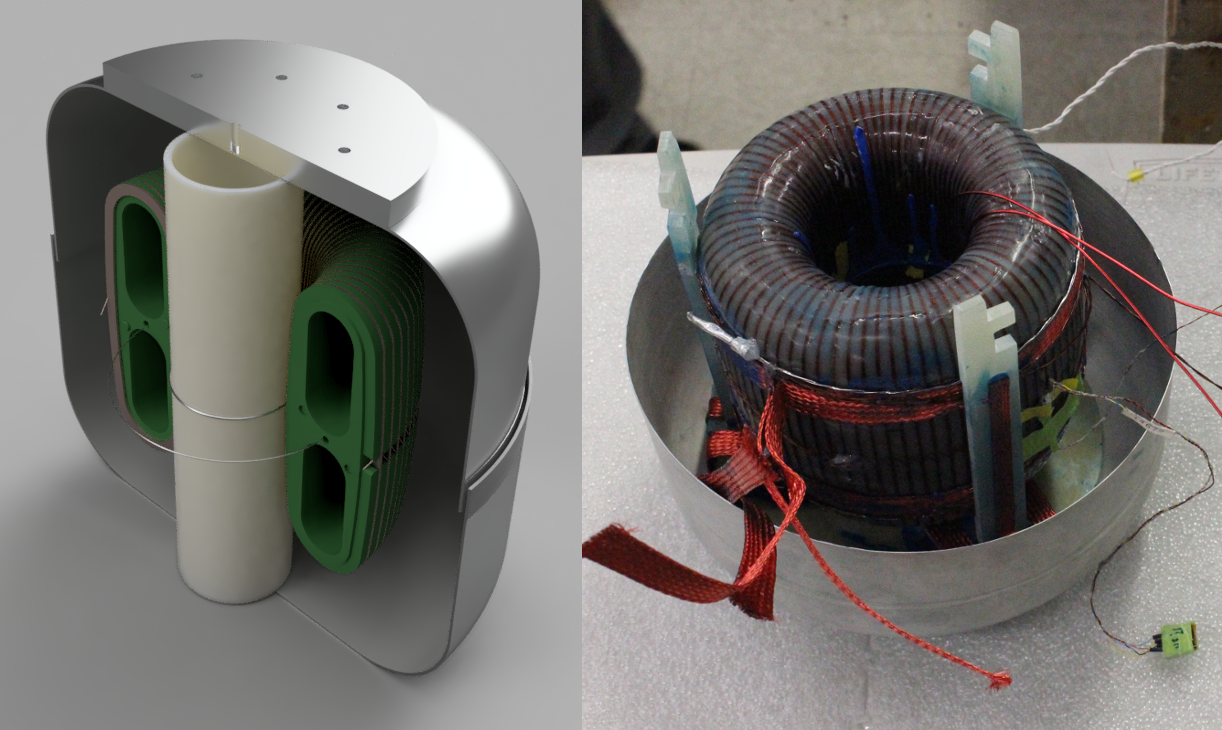}
  \caption{\emph{Left:} Rendering of the \abra setup. The primary magnetic field is driven by 1,280 superconducting windings around a \POM support frame (\emph{green}). The axion-induced field is measured by a superconducting pickup loop mounted on a PTFE support (white). A second superconducting loop runs through the volume of the magnet to produce a calibration signal. All of this is mounted inside a superconducting shield. \emph{Right:} Picture of the exposed toroid during assembly.}
    \label{fig:ABRAPhoto}
\end{figure}

\abra consists of a superconducting persistent toroidal magnet produced by Superconducting Systems Inc. \cite{ssi} with a minimum inner radius of 3\,cm, a maximum outer radius of 6\,cm, and a maximum height of 12\,cm. The toroidal magnet is counter-wound to cancel azimuthal currents; see \cite{ABRA_10cm_Technical} for details. We operate the magnet in a persistent field mode with a current of 121\,A, producing a maximum field of 1\,T at the inner radius. We confirmed this field with a Hall sensor to a precision of $\sim1$\,\%. Due to the toroidal geometry of the magnet, the field in the center should be close to zero (in the absence of an axion signal). 

To reduce AC magnetic field noise, we use both magnetic shielding and vibrational isolation. The toroid is mounted in a G10 support inside a tin-coated copper shell which acts as a  magnetic shield below 3.7\,K, when the tin coating becomes superconducting. The toroid/shield assembly is thermalized to the coldest stage of an Oxford Instruments Triton 400 dilution refrigerator and cooled to an operating temperature of $\sim1.2$\,K. The weight of the shield and magnet is supported by a Kevlar string which runs $\sim$2\,m to a spring attached to the top of the cryostat. This reduces the mechanical coupling and vibration between the detector and cryostat.

We measure AC magnetic flux in the center of the toroid with a solid NbTi superconducting pickup loop of radius 2.0\,cm and wire diameter 1\,mm. The induced current on this pickup loop is carried away from the magnet through $\sim50$\,cm of 75\,$\mu$m solid NbTi twisted pair readout wire up to a Magnicon two-stage SQUID current sensor. 
The 75\,$\mu$m wire is shielded by superconducting lead produced according to \cite{Solder-Paper}. The majority of the 1\,mm wire is inside the superconducting shielding of the magnet, but about 15\,cm is only shielded by stainless steel mesh sleeve outside the shield. 

The two-stage Magnicon SQUID current sensor is optimized for operation at $<1$\,K; we operate it at $870$\,mK. The input inductance of the SQUID is $L_{\rm in}\approx150$\,nH and the inductance of the pickup loop is $L_p\approx100$\,nH. The SQUID is operated with a \FLL to linearize the output, which limits the signal bandwidth to $\approx 6$\,MHz. We read out the signal with an AlazarTech ATS9870 8-bit digitizer, covering a voltage range of $\pm40$\,mV. The digitizer is clocked to a Stanford Research Systems FS725 Rb frequency standard. In order to fit the signal into the range of our digitizer, we filter the signal through a 10\,kHz high-pass filter and a 1.9\,MHz anti-aliasing filter before sending it to the digitizer.

To calibrate the detector, we run a superconducting wire through the volume of the toroid at a radius of 4.5\,cm into which we can inject an AC current to generate a field in the pickup loop, similar to what we expect from an axion signal. The coupling between the calibration and pickup loop can be calculated from geometry to be $\approx$50\,nH. We perform a calibration scan to calculate the end-to-end gain of our readout system. Our calibration measurements indicate that our pickup-loop flux-to-current gain is lower than expected by a factor of $\sim6$. We determined this to be likely due to parasitic impedances in the circuit, and we will address this issue in future designs.

\begin{figure*}
  \centering
  \includegraphics[width=\textwidth]{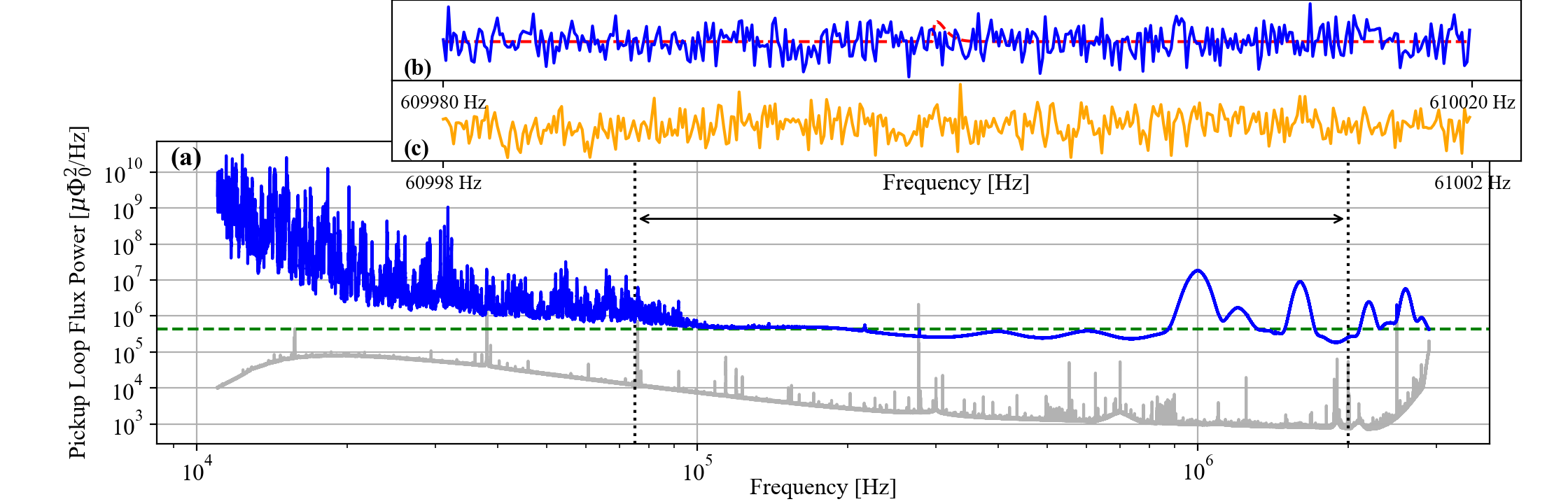}
  \caption{Flux spectrum averaged over the the data used in this analysis. (\emph{a}) The spectrum over the frequency range \mbox{$11\,\mathrm{kHz}<f<3\,\mathrm{MHz}$}, corrected for the pre-digitizer filters (\emph{blue}). For comparison, we also show the digitizer noise floor, corrected for pre-digitizer filters (\emph{gray}) and the characteristic SQUID flux floor (\emph{green dashed}). The axion search range is between the dotted black lines. (\emph{b}) A zoomed view of the 10\,MS/s spectrum (\emph{blue}) with $\Delta f = 100$\,mHz and and an example axion signal at the 95\% upper limit (\emph{red dashed}). (\emph{c}) A zoomed view of the 1\,MS/s spectrum with $\Delta f=10$\,mHz. Note that the digitizer data was collected at a different time from the SQUID data, and shows a few transient peaks that are not present in the SQUID data.
 }
    \label{fig:spectrum}
\end{figure*}

\section{Data Collection}
We collected data from July 16, 2018 to August 14, 2018, for a total integration time of $T_{\rm int}=2.45\times10^{6}$\,s. The data stream was continuously sampled at a sampling frequency of 10\,MS/s for the duration of the data-taking period. After completing the magnet-on data run, we collected two weeks of data with the magnet off, but otherwise in the same configuration.

During data taking, the data follow two paths. First, we take the \DFT of individual sequential 10\,s buffers of $10^8$ samples each to produce a series of {\PSD}s. These are accumulated together to produce an average \PSD, called \Pten, with a Nyquist frequency of 5\,MHz and a frequency resolution of $\Delta f=(10\,\mathrm{s})^{-1}=100$\,mHz. In the second path, the streamed data are decimated by a factor of 10, to a sampling frequency of 1\,MS/s, collected into a 100\,s buffer of $10^8$ samples, then transformed and compiled into a similar running average \PSD, \Pone, with Nyquist frequency of 500\,kHz and $\Delta f=(100\,\mathrm{s})^{-1}=10$\,mHz. The 1\,MS/s data stream is further decimated in real time to a 100\,kS/s stream and written directly to disk. This can be transformed offline to produce \Phun, with a Nyquist frequency of 50\,kHz and $\Delta f=1/T_{\rm int}\approx 408$\,nHz. We do not use \Phun for the present search. All \DFT transforms are taken with the FFTW3 library \cite{FFTW05}. 

The \Pten spectra are written to disk and reset after every 80 averages; each stored spectrum thus covers a period of 800\,s. This allows us to separate time-dependent noise signals from a constant axion signal. Similarly, the \Pone spectra are written to disk and reset every 16 averages, and cover a period of 1600\,s.  Figure~\ref{fig:spectrum} shows the full \Pten spectrum as well as close-ups of the \Pone spectra, converted to pickup loop flux spectral density using the calibration measurements. 

Each of the \Pten, \Pone and \Phun spectra have a usable range limited by the Nyquist frequency on the high end, and the frequency resolution required to resolve a potential axion signal on the low end. With our sampling frequency and integration times, we could perform a search over the range from 440\,mHz -- 5\,MHz with enough resolution that a potential signal would span 5 -- 50 frequency bins (assuming a typical ADM velocity of $\sim220$\,km/s), though in practice our search range is limited by the signal filters. 

We observed large $1/f$-type behavior below $\sim$20\,kHz, with broad noise peaks extending up to $\sim$100\,kHz. This noise is strongly correlated with vibration on the top plate of the cryostat up to the highest frequency measured by our accelerometer, $\sim10$\,kHz \cite{ABRA_10cm_Technical}. We believe that the tail of this noise continues up to higher frequencies before becoming sub-dominant to the flux noise of the SQUID above 100\,kHz. This noise degrades our sensitivity at lower frequencies and we restrict our search range to \mbox{$75\,\mathrm{kHz} < f < 2\,\mathrm{MHz}$}.

For $\sim$1 week after starting the data collection, we observed very narrow and variable noise peaks in the \PSD above $\sim1.2$\,MHz. We are investigating the source of these peaks. After about a week, these peaks died away slowly and did not return until we re-entered the lab to refill an LN$_2$ dewar, then died away again after a few days. The affected time periods were removed and account for a $\sim$30\% decrease in our exposure. We hope that in the future, a more detailed analysis will allow us to recover a significant fraction of this lost exposure.

\section{Data Analysis Approach}

Our data analysis procedure closely follows the method introduced in \cite{Foster2018}. Our expected signal is a narrow peak in the pickup loop \PSD, with a width $\Delta f/f\sim10^{-6}$ arising from the DM velocity dispersion. When averaged over $N_{\rm avg}$ independent PSDs, the signal in each frequency bin $k$ ($f_k$) will follow an Erlang distribution with shape parameter $N_{\rm avg}$ and mean
\begin{eqnarray}
s_k = 
\left\{ \begin{array}{lc}
A \left. \frac{\pi f(v)}{m_a v} \right|_{v = \sqrt{4\pi f_k/m_a - 2}} & f_k > m_a/2\pi\,, \\
0 & f_k \leq m_a/2\pi\,,
\end{array} \right.
\end{eqnarray}
where $A$ is defined in Eq.~(\ref{eq:pickup_flux_power}). We assume $f(v)$ is given by the Standard Halo Model, with velocity dispersion $v_0 = 220$\,km/s/$c$, and $v_{\rm obs} = 232$\,km/s/$c$ the DM velocity in the Earth frame \cite{McMillan:2009yr}, with $c$ the speed of light. With the DM density and velocity distribution specified, the only free parameter in the predicted signal rate is \gagg.

We expect our background noise sources to be normally distributed in the time domain, such that when combined with an axion spectrum, the resulting \PSD data is still Erlang-distributed.  Accordingly, our combined signal-plus-background model prediction in each frequency bin is an Erlang distribution $P(\bar{\mathcal{F}}_k; N_{\rm avg}, \mu_k)$ with shape parameter $N_{\rm avg}$ and mean $\mu_k = s_k + b$ (see \cite{Foster2018} for details). Although the background \PSD varies slowly with frequency, the axion signal for a given mass is narrow enough that we restrict to a small frequency range and parameterize the background as a constant $b$ across the window. We verified that the results of our analysis were not sensitive to the size of the window chosen.

We performed our analysis on the \Pten and \Pone spectra over frequency ranges 500\,kHz to 2\,MHz and 75\,kHz to 500\,kHz, respectively. We chose the frequency at which we transition from one set of spectra to the other so that the axion signal window is sufficiently resolved everywhere, though we have seen that the exact choice has little effect on the final result.   
We rebin the \Pten (\Pone) spectra in time into 53 (24) spectra that cover 32,000\,s (64,000\,s) each. This was done to speed up processing time, though it is not necessary for our analysis approach.

\begin{figure*}[htb]
\centering
\includegraphics[width=1.0\textwidth]{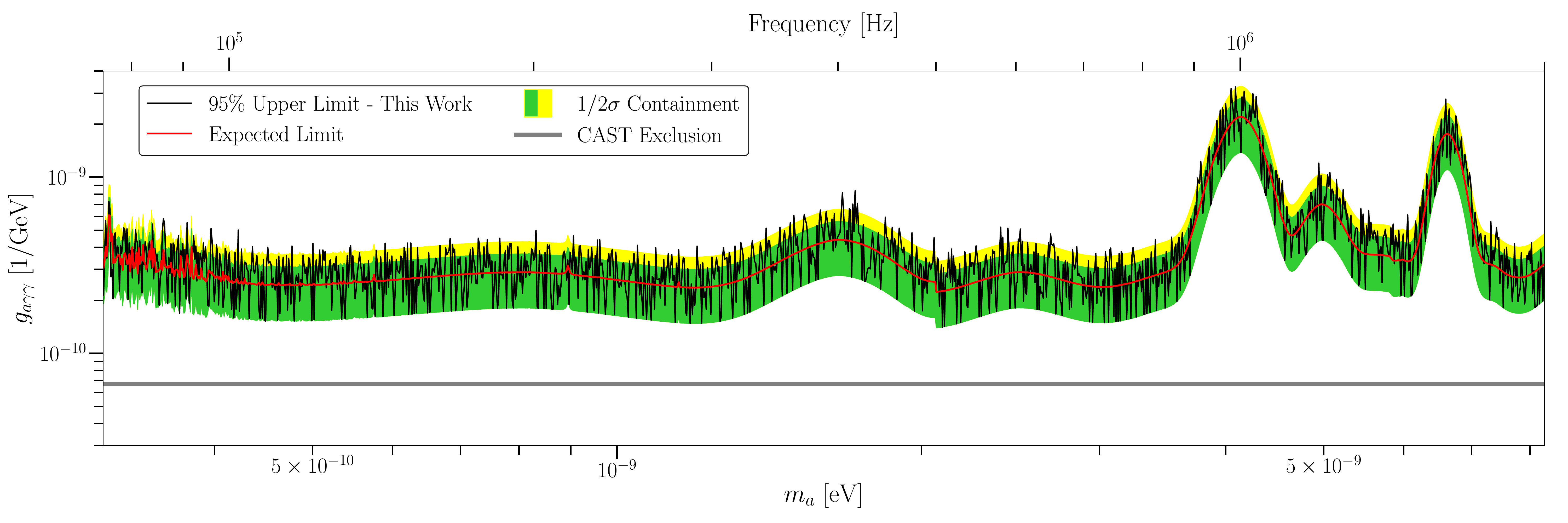}
\caption{ The limit on the axion-photon coupling $g_{a \gamma \gamma}$ constructed from ABRACADABRA-10cm data described in this work.  We compare the observed limit, which has been down-sampled in the number of mass points by a factor of $10^4$ for clarity of presentation, to the expectation for the power-constrained limit under the null hypothesis. This down-sampling excludes the 87 isolated mass points vetoed in the discovery analysis; further details will be presented in \cite{ABRA_10cm_Technical}.  Additionally, we show the astrophysical constraint on $g_{a\gamma\gamma}$ in this mass range from the CAST helioscope experiment \cite{CAST2017}; the region above the grey line is excluded. }
\label{fig:ExclusionCurve}
\end{figure*}

We test for an axion signal at mass $m_a$ and coupling strength $A$ by constructing a joint likelihood of Erlang distributions over the 53 (24) \Pten (\Pone) given the observed \PSD data~\cite{Foster2018,ABRA_10cm_Technical}. For each axion mass, we assign a unique background nuisance parameter to each of the rebinned \Pten (\Pone) spectra and profile over the joint likelihood to construct the profile likelihood for $A$ at that mass.  This accounts for the possibility that the background level might change on timescales of hours to days.

To detect an axion signal, we place a 5$\sigma$ threshold on a discovery \TS. To evaluate this we first calculate the profile likelihood ratio $\lambda(m_a,A)$, at fixed $m_a$, as the ratio of the background-profiled likelihood function as a function of $A$ to the likelihood function evaluated at the best-fit value $\hat A$. From here, we define $\text{TS}(m_a) = -2 \log \lambda(m_a,0)$ for $\hat A > 0$ and zero otherwise. This quantifies the level at which we can reject the null hypothesis of $A=0$. The $5\sigma$ condition for discovery at a given $m_a$ is ${\rm TS}(m_a) > {\rm TS}_{\rm thresh}$, where \cite{Foster2018}
\begin{equation}
{\rm TS}_{\rm thresh} = \left [ \Phi^{-1}\left(1 - \frac{2.87 \times 10^{-7}}{N_{m_a}}\right) \right]^2
\end{equation}
accounts for the local significance as well as the \LEE for the $N_{m_a}$ independent masses in the analysis (here $\Phi$ is the cumulative distribution function for the normal distribution with zero mean and unit variance). For this analysis, $N_{m_a} \approx8.1\times10^6$ between 75\,kHz and 2\,MHz, and ${\rm TS}_{\rm thresh} = 56.1$.

Where we have no detection, we set a 95\%~C.L.\ limit, $A_{95\%}$, again with the profile likelihood ratio. To do so, we use the statistic $t(m_a,A) = -2 \log \lambda(m_a,A)$, with $A > \hat A$, by $t(m_a,A_{95\%}) = 2.71$.  We implement one-sided power-constrained limits~\cite{Cowan:2011an}, which in practice means that we do not allow ourselves to set a limit stronger than the $1\sigma$ lower level of the expected sensitivity band. We compute the expected sensitivity bands using the null-hypothesis model and following the procedure outlined in~\cite{Foster2018}.

We had to exclude a few specific mass points from our discovery analysis due to narrow background lines that were also observed when the magnet was off. To veto these mass points as potential discoveries, we analyze data collected while the magnet was off (where no axion signal is expected) using the same analysis framework.  If in this analysis we find a mass point with {\LEE}-corrected significance greater than 5$\sigma$, we exclude that mass point from our axion discovery analysis. In total, this procedure ensures a signal efficiency of $\gtrsim99.8$\%. Our axion search yielded 83(0) excesses with significance $\geq5\sigma$ in the frequency range 500\,kHz to 2\,MHz (75\,kHz to 500\,kHz), however all of these points are vetoed by the magnet-off data. We do not exclude these points from our upper-limit analysis, though the observed limits at these isolated points are weaker (they do not appear in Fig.~\ref{fig:ExclusionCurve} because of down-sampling for clarity).

We verified our analysis framework by injecting a simulated software axion signal into our real data and confirmed that the data-quality cuts and analysis framework described above are able to correctly detect or exclude the presence of an axion signal. In the future we hope to build this into a hardware-based option, using the calibration loop to inject ``blinded'' signals similar to the approach used by ADMX \cite{ADMX2018}. Further details of the analysis and statistical tests we have performed, as well as an extended discussion of the noise in the excluded exposure, will be further described in a future publication \cite{ABRA_10cm_Technical}.
 
\section{Results and Discussion}

We observe no evidence of an axion signal in the mass range $3.1\times10^{-10}$\,eV -- $8.3\times10^{-9}$\,eV and place upper limits on the axion-photon coupling $\gagg$ of at least \mbox{$3.3\times10^{-9}$\,GeV$^{-1}$} over the full mass range and down to \mbox{$1.4\times10^{-10}$\,GeV$^{-1}$} at the strongest point. Our full exclusion limits are shown in Fig.~\ref{fig:ExclusionCurve}. This result represents the first search for \ADM with $m_a<1\,\mu$eV, and with one month of data is already competitive with the strongest present astrophysical limits from the much larger CAST helioscope \cite{CAST2017} in the range of overlap. 

We note that for a significant range in frequency, we achieved the SQUID noise-limit. However, constraints on the detector configuration introduced parasitic impedances into the readout circuit, which lead to a loss in the ultimate axion coupling sensitivity \cite{ABRA_10cm_Technical}. This will be addressed in future efforts and could yield up to a factor of $\sim$6 improvement in sensitivity with a similar exposure.

As \abra is a prototype detector, there are many potential directions for future improvement. Our focus in this work has been on demonstrating the feasibility and power of this new approach. Future upgrade paths for the ABRACADABRA program will include improvements to shielding and mechanical vibration isolation, reduction of parasitic inductances, improvements to the readout configuration, expanded frequency range, and construction of a larger toroid.

\section{Conclusion}
The axion is a promising DM candidate, but its couplings remain largely unconstrained at low masses. In this Letter we have demonstrated the capabilities of the broadband axion search \abra, which can cover many decades of axion mass with relatively short data-taking time and which has nonetheless set competitive limits on the coupling $\gagg$. We have already identified stray fields from the toroid, vibration, and cross-talk with the DAQ to be significant sources of background \cite{ABRA_10cm_Technical}. Understanding these sources will be critical for scaling the experiment up to the $\sim 1-5$\,m scale to search for the QCD axion.

The \ABRA\ program is highly complementary to microwave cavity experiments like ADMX and HAYSTAC, as well as proposed experiments like MADMAX \cite{TheMADMAXWorkingGroup:2016hpc}, which can probe the coupling $\gagg$ for axion masses in the range $10^{-6}-10^{-5}\,\eV$ and $10^{-5}-10^{-3}\,\eV$, respectively. By demonstrating the efficacy of \abra, we have set the stage for a full-scale experiment which can probe the QCD axion \cite{ABRA_10cm_Technical}. The combination of \ABRA, microwave cavities, and MADMAX type detectors, along with experiments probing the axion-nucleon coupling \cite{PhysRevX.4.021030} and indirect constraints from black hole superradiance \cite{Arvanitaki:2014wva,Arvanitaki:2016qwi}, will be able to probe the range of couplings expected for typical QCD axion models in the coming decades.

\begin{acknowledgments}
\it
The authors would like to acknowledge the useful conversations and advice from Henry Barthelmess of Magnicon Inc. This research was supported by the National Science Foundation under grant numbers NSF-PHY-1658693, NSF-PHY-1806440, NSF-PHY-1505858 and NSF-PHY-1122374. This work was also supported by DOE Early Career Grant number DE-SC0019225, by the Kavli Institute for Cosmological Physics at the University of Chicago, by the Miller Institute for Basic Research in Science at the University of California, Berkeley, by the MIT Undergraduate Research Opportunities in Physics program, by the Leinweber Graduate Fellowship Program and by the Simons Foundation through a Simons Fellowship in Theoretical Physics. This research was supported in part through computational resources and services provided by Advanced Research Computing at the University of Michigan, Ann Arbor. This material is based upon work supported by the U.S. Department of Energy, Office of Science, Office of Nuclear Physics under Award Numbers DE-FG02-97ER41041 and DEFG02-97ER41033 and by the Office of High Energy Physics under grant DE-SC-0012567. We would like to thank the University of North Carolina at Chapel Hill and the Research Computing group for providing computational resources and support that have contributed to these research results.
\end{acknowledgments}

\clearpage

\end{document}